# Phase transition of a single star polymer: a Wang-Landau sampling study

Zilu Wang[1] and Xuehao He*[2]

[1] Department of Polymer Science and Engineering, School of Chemical Engineering and Technology, Tianjin University, 300072 Tianjin, China;

[2] Department of Chemistry, School of Science, Tianjin University, 300072 Tianjin, China.

**Abstract:** Star polymers, as an important class of nonlinear macromolecules, process special thermodynamic properties for the existence of a common connecting point. The thermodynamic transitions of a single star polymer are systematically studied with the Bond Fluctuation Model using Wang-Landau sampling techniques. A new analysis method employing the shape factor is proposed to locate the coil-globule (CG) and liquid-crystal (LC) transitions, which shows a higher efficiency and accuracy than the canonical specific heat function. The LC transition temperature is found to obey the identical scaling law as the linear polymers. However, the CG transition temperature shifts towards the LC transition with the increasing of the arm number. The reason is that for the star polymer a lower temperature is needed for the attractive force to overcome the excluded volume effect of the polymer chain because of its high arm density. This work clearly proves the structural distinction of the linear and star polymers can only affect the CG transition while has no influence on the LC transition.

**Keywords**: star polymers, Wang-Landau sampling, coil-globule transition, liquid-crystal transition.

*Corresponding author, E-mail address: xhhe@tju.edu.cn

# I. INTRODUCTION

The phase transition behavior of a single polymer chain is a fundamental topic in polymer science. Compared with small molecules, polymers show a more complex thermodynamic property such as coil-globule-crystal transitions due to the bonding linkages between monomers, which attracted more attentions. Recent years, the developments of nonlinear polymers, such as tree-like,[1] ladder-like,[2] comb-like,[3] and ring-shaped[4] polymers, have aroused growing interests of material scientists because of their complex structures and unique properties for advanced material applications.[5] Among them, star polymers can be regarded as the simplest branched-type polymers where multiple linear chains connected to a common core. Their regular structures and tunable interactions between arms make them becoming suitable models to probe the general properties of more complex polymer architectures experimentally.[6] Using star polymers, material scientists have developed self-assembly building blocks,[7-10] linear polymer modifiers[11] and various functional materials.[6,12] Although star polymers own such widely applications, the theoretical descriptions[13-18] of their thermodynamic transitions are still rare compared with linear polymers. In the past few decades, the thermodynamic transitions of linear polymers had been well studied through both theoretical calculations and simulations.[19-36] The phase transition of a polymer chain usually goes accompanying with the collapse of the whole macromolecule when they are in a poor solvent or the temperature is changed. During the collapse of a linear polymer chain, three types of morphologies can be observed, i.e. expanded coil (the gaseous state), compact globule (the liquid state) and regular crystal (the solid state). For the temperature-induced collapse, linear polymers usually undergo a two-stage transition, i.e. a $2^{nd}$ order coil-globule (CG) transition at the higher temperature and a 1$^{st}$ order liquid-crystal (LC) transition at the lower temperature.

Recent years, the remarkable advances in statistical studies with both lattice[24,27-29,34-36] and off-lattice[20-23,30,32] polymer models had deeply touched the detailed mechanisms of these transitions. Accurate information has been obtained through many advanced simulation techniques. For flexible linear polymers, the transition behaviors can be altered both by the size of the polymer chain and the type/range of the interaction between monomers. The two typical transitions mentioned above may separate or combine when the inter-monomer interactions changed.[22,23] Therefore, the design of the interaction between monomers may provide a feasible way to obtain the required thermodynamic properties. More importantly, the general knowledge about the thermodynamic transitions of linear polymers is helpful to understand the transition phenomena in bio-macromolecules, such as the nucleic acid helix[37] and protein folding problems.[38] These studies about the transitions of linear polymers enlighten us to think about whether the topological structure of a polymer can also affect its thermodynamic transition behavior. Compared with designing the interaction between monomers, polymer structure is another important source to tune their properties. However, the influences of the polymer structures (linear / nonlinear) on the transition properties are still not fully clear.

In this study, the phase transitions of a single star polymer are investigated with the lattice polymer model. The dependence of the structural factors (i.e. the arm length and arm number) on its transition behavior is explored. The description of the model and the implementation of the sampling method for the star polymer are presented in section II and III, respectively. In section

IV, the results of the thermodynamic transitions and structural properties of the star polymer are analyzed and discussed. The conclusion remarks are given at the final part.

## II. MODEL

The star polymer is constructed with the 3D cubic lattice Bond Fluctuation Model (BFM),[39] where each monomer occupies eight lattice positions and the bond length between two linked monomers can fluctuate from 2 to $\sqrt{10}$ (totally five different bond lengths are allowed, detailed algorithms see Binder *et al.*[39]). The flexible bonding rules within BFM allow one to build various types of polymers, such as linear polymers,[23,25,28,29,33] ring polymers,[40,41] comb/brush-like polymers,[42] star polymers,[43] hyper-branched polymers[44,45] or even dendritic polymers.[46] In this study, the "core first" method is adopted to create the star polymer. Firstly, the core monomer is placed at the center of a cubic box (large enough to avoid the periodic boundary effect), and then its arms grow up towards different directions. The arm number ($N_a$) in the present work ranges from 2 to 12, and the arms in one star polymer have the same length ($L_a$) ranging from 3 to 100. The snapshots of an expanded coil and a collapsed crystallite for the linear polymer ($N_a$=2 and $L_a$=240) and the 12-arm star polymer ($N_a$=12 and $L_a$=60) are shown in Fig. 1. The crystallite has a much smaller radius than its expanded coil during the collapse transition.

In this work, the collapse of the polymer is assumed to be induced by temperature and the thermal related properties are derived from its energy. The Hamiltonian of this polymer model is identical to Rampf *et al.*'s[28] with the form:

$$H = -\frac{1}{2}\sum_{i}\sum_{j\neq i}U(r_{ij}) \qquad U(r_{ij})=\begin{cases}\varepsilon & r_{ij}\leq\sqrt{6}\\ 0 & r_{ij}>\sqrt{6}\end{cases} \qquad (1)$$

The interaction between the adjacent monomers is a square well like attractive potential (the well depth is $\varepsilon$) which is given by Eq. 1. This potential is calculated between all monomer pairs including the bonded neighbors. Here, the depth of the well is set as $\varepsilon=1$ (lattice unit), therefore the potential of the polymer is calculated by counting the number of the nearest neighboring pairs. In the present work, the interaction between monomers is short-range dominated that the effective range is within $\sqrt{6}$, (i.e. the radius of effective interaction $r_{ij}=2, \sqrt{5}$ and $\sqrt{6}$). Past studies applying this model[25,28,29] and similar investigations with off-lattice model [21-23] all suggested that the interaction range can affect the transition behaviors of linear polymers, where a single chain polymer with short-range interaction shows a direct discontinuous coil-crystal transition without the globule state. Different from these studies, our work focused on the structure of polymers instead of the interaction types. With the Hamiltonian defined as Eq. 1, the two transitions are observable for linear polymers under the finite size condition.[34] However, it is noted that, in the limit of the chain length $N \to \infty$, there is only one direct transition from the coil to crystal under this potential. The focusing point of this study is the influence of the polymer structure on its transitions under the finite size condition.

**Movement algorithm for the star polymer chain**

The Wang-Landau (WL) sampling algorithm[47,48] is employed to calculate the relative density of state of the star polymer. It should be noticed that the efficiency of the WL algorithm strongly depends on the conformation movement algorithm. In this work, a hybrid movement algorithm, which combines four types of simple movements, is used to speed up the sampling. Figure 2 shows the illustration of the four movement algorithms[31], which are consist of the BFM movement (randomly hop one monomer towards different directions), the pivot movement (randomly rotate a chain according to a pivot bond), the crankshaft movement (randomly rotate a segment of the polymer chain), and the center imaging movement (consist of three steps: (1) randomly select two units to determine a segment of the polymer chain; (2) generate a new image conformation according to the center of the two units. (3) reverse the numbers of the units in this segment). For linear polymers,[34] the slithering snake algorithm provides an efficient way to sample low temperature state, which, however, cannot be applied to the star or other nonlinear polymers since it will break up the topological structure. These four movement algorithms change the conformation in different ways. The short-range BFM movement relaxes the polymer chain by varying the bond length and gives minor changes to the conformation. The other three long-range movements generate larger changes on the conformation to overcome the large entropy barriers. This hybrid algorithm provides an effective way to traverse conformational space with ergodic condition and it should be applicable to all nonlinear polymer chains with lower branching degree. In our study, the ratio of these four types of movements is every 100 times of the BFM movements with one time of the other three movements, which shows a well sampling speed.

## III. METHOD

**a) Wang-Landau sampling**

For sampling rare events such as the phase transitions, the Wang-Landau sampling algorithm[47,48] enables the sampler running uniformly in the energy space to across the entropy barrier, and bridges the state tunnels that traditional canonical sampling method may never reach. The accurate

1  estimation of the micro-canonical density of state (DOS) function $g(E)$ is also a feature that the

2  WL algorithm is better than the traditional sampling algorithms. When performing the WL

3  sampling, the acceptance criterion is temperature independent and adopted as:

$$p(E_{old} \Rightarrow E_{new}) = \min(1, g(E_{old})/g(E_{new})) \qquad (2)$$

5  The above acceptance manner generates a driving force to the sampler to walk over the whole

6  energy space through proper Monte Carlo movement algorithms to make the energy state

7  histogram *"flat"*. The energy space is divided into many discrete bins representing the energy

8  states. With the sampling carrying on, the sampled conformations are recorded at the

9  corresponding energy bin into a histogram array $H(E)$. The histogram array and the DOS array

10 are initialed as $H(E)=0$ and $g(E)=1$ for all $E$. The first sampled conformation is

11 intentionally accepted, and for the second and later conformations, the acceptances are judged via

12 Eq. 2. If a sample with the energy state transition $(E \Rightarrow E')$ is accepted, then

13 $H(E')+1 \Rightarrow H(E')$ and $g(E') \times f \Rightarrow g(E')$, if rejected, then $H(E)+1 \Rightarrow H(E)$ and

14 $g(E) \times f \Rightarrow g(E)$. When the number of the sampled conformations is sufficiently large or some

15 conditions (such as flatness>90%) are satisfied, the updater $f$ decreases with a certain manner.

16 With the sampling going on, $f$ gradually approaches to 1, and the $g(E)$ approaches to the true

17 DOS. Usually, $\ln g(E)$ and $\ln f$ are used in sampling to avoid the numerical overflow since

18 the DOS is often very large. Correspondingly, the updating manner of $\ln g(E)$ is rewritten as

19 $\ln g(E) + \ln f \Rightarrow \ln g(E)$. In micro-canonical ensemble[49], the entropy function is defined as:

$$S(E) = k_B \ln g(E) \qquad (3)$$

21 where $k_B$ is Boltzmann constant, which is reduced to 1 in our model. Using the logarithm of DOS

function, the micro-canonical entropy function can be calculated directly from the WL algorithm.

In the original version of WL algorithm, the decaying manner of the updater $f$ is exponentially as $\ln f \Leftarrow \ln f / 2$. However, this updating manner had been proved to be not sufficiently converged since the decaying speed of the updater $f$ is too fast compared with the accumulation of the samples, which leads to the error saturation problem.[50] Recent studies showed that the convergence of the WL method is related to the simulation time $t$, which is defined as $M/N_E$ (where $M$ is the Monte Carlo Steps of the simulation, $N_E$ the number of energy states) in the original literature of $1/t$ WL algorithm.[50] The classical 2D Ising model and numerical integration benchmarks[50,51] had shown that the $1/t$ WL algorithm provides a correct error convergence balance. With the decreasing of the updater $\ln f$, the $1/t$ algorithm enables the statistical error evolving as $\sqrt{\ln f}$.[50] The $1/t$ WL algorithm indeed makes a progress in deeply understanding the mechanism of the WL algorithm, however, a very recent study[52] on the convergence of the $1/t$ algorithm in sampling polymer model shows that the convergence of the $1/t$ also fails to obtain the accurate DOS function due to the complex energy landscape. In this study, another definition of the simulation time $t$ = ergodic times (when all energy state is visited, call it one time ergodicity) is adopted to complement the original $1/t$ WL sampling algorithm.[50,51,53] This modification may lead to a longer simulation time to sample very complex system, however, it is necessary to wait a longer time for a slower sampler to obtain the accurate information from sampling. This is why the conformation movement algorithms are so important to enhance the efficiency of the WL algorithm. The value of the updater is crucial to the accuracy of the results, which determines the fluctuation of the DOS function. Previous studies had shown that,[50,51,53,54] with a fixed $\ln f$, the final fluctuation or the statistical error is within the $\sqrt{\ln f}$. The modified $1/t$ WL method also have the statistical error $\sqrt{\ln f}$,[50,55] but the final convergence is very slow. To find a good balance between accuracy and computational cost, we set the initial updater $\ln f_0 = 10^{-3}$ to accumulate enough samples and the ergodic time is set as $10^4$ for refining the DOS function, which makes the final updater $\ln f = 10^{-7}$, therefore, the statistical error is $\delta \approx 0.0003$ for the DOS functions within our work.

**b) Parallel sampling**

This version of $1/t$ WL algorithm is slower than the original one due to the huge amount of calculation. To obtain meaningful information within an acceptable time scale, a parallel version of the WL sampling algorithm[56] is used to accelerate the sampling speed. In this algorithm, multiple sampling threads are created (using the OPENMP technique) to sample the whole range of energy space instead of dividing it into several windows used in the original WL algorithm. All sampling threads share and update the only common $\ln g(E)$ array with the identical updater $\ln f$. In our study, the atomic operation is not adopted to modify the $\ln g(E)$ array since it has no influence on the sampling results. This procedure eliminates the errors caused by the interspace of the overlapped energy windows and accelerates the sampling speed greatly (linear speedup proportional to the threads number can be obtained from our tests). On modern multi-cores and multi-ways computer, this parallel algorithm reduces much calculation time compared with single thread sampling method.

## IV. RESULTS and DISCUSSIONS

**(1) Density of state**

The DOS function $g(E)$ is the distribution density of the conformations within the energy range: $[E_{min}, 0]$. The determination of $E_{min}$ is important to the final calculation accuracy and computational cost. A tiny extension of the $E_{min}$ value may need huge amounts of calculation time. Therefore, a balance between the calculation accuracy and computation cost requires a careful choice. Past study[34] had shown the transition behavior can be observed when the energy range reaches $E_{min} = -4N \sim -6N$. In our calculation, the possible energy range ($[E_{min}, 0]$) is determined by a longtime test run for small polymer chains $(N < 30)$. While for larger ones $(N \geq 30)$, we set the lowest energy value $E_{min}$ to -4N (same as Rampf et al.[34]).

The dependence of the entropy functions $\ln g(E)$ on the specific energy $E/N$ are shown in Fig. 3(a1~c1) for three types of polymers. These three types of polymers, i.e. the linear polymers ($N_a=2$), the 6-arm star polymers ($N_a=6$) and the 12-arm star polymers ($N_a=12$), show a similar shape in their entropy curves. Here, the entropy functions are aligned at the point $E=0$ to satisfy the condition $\ln g(0) = 0$ (i.e. $\ln g(E) \Leftarrow \ln g(E) - \ln g(0)$). With the increasing of the number of monomers $N$, the widening of the entropy function between the highest and lowest points indicates the expanding of the conformation space. Particularly, in Fig. 3(a1~c1), a small intersection region can always be observed in each panel. The intersection region looks like a fixed point of the entropy function that is unrelated with number of monomers. Since the entropy function is aligned according to the highest energy state, it is believed that this intersection region represents a balance point, similar to the Gaussian coil state in the polymer conformations, which is size independent. However, it is noted that these intersection regions shift towards the higher entropy direction with the increasing of the arm number $N_a$. Figure 4 plots the linear relationship of the $\ln g(E)$ value of these intersection region as a function of the arm number cubed $N_a^3$. This special linear relation may represent the stretching dimension of their arms. The average entropy functions $\ln g(E)/N$ are also shown in Fig. 3(a2~c2) and the changing trends for the three types of polymers are clearly labeled (see arrows). For linear polymers (Fig. 3(a2)), the average entropy increases with the growing of the arm length $L_a$ and the maximum average entropy shifts towards the lower energy direction and these curves finally tend to a fixed shape. For star polymers ($N_a$ =12 in Fig. 3(c2)), however, the average entropy decreases with the increasing of the arm length $L_a$ and the maximum average entropy shifts towards the higher energy direction. The average entropy function corresponds to the average number of conformations permitted at certain energy state. With the growth of linear polymer chains, the conformation number of linear polymers reasonably increases due to the expansion of the average free space and results in the increasing of the average entropy. Meanwhile, the energy state value at the maximum average entropy decreases to overcome the overly stretching of the short linear polymer at higher energy state. However, the situations are different for star polymers, which have a higher monomer density for the existence of the center core. With the growing of their arms, the conformation number of star polymers at higher energy state inevitably decreases due to the too high monomer density, which leads to the decrease of the average entropies at higher energy state. While the increase in the energy state value at the maximum average entropy is due to the relatively decrease of the average monomer density. It indicates that, under the finite size condition of polymers, the structural difference between linear polymers and star polymers strongly affect their density of state functions.

**(2) Canonical analysis**

The thermodynamic transition of a single polymer is under the canonical ensemble with the temperature $T$. The corresponding energy distribution of the system can be described as the Boltzmann-Gibbs distribution. In this section, the canonical analysis method is utilized for the phase transition of the single star polymer.

In this extended ensemble, the partition function is expressed as:

$$Z(T) = \sum_E g(E) e^{-E/k_B T} \qquad (4)$$

By averaging over all quantities within the canonical ensemble, any thermal or structural quantity can be calculated from the DOS functions. Equation 5 shows the general formula to calculate the dependence of any quantities on temperature.

$$\langle Q \rangle_T = \sum_E \langle Q \rangle_E P(E,T) \qquad (5)$$

$$P(E,T) = g(E) e^{-E/k_B T} / Z(T) \qquad (6)$$

where $P(E,T)$ is the canonical probability function, $\langle Q \rangle_T$ the average quantity at $T$, $\langle Q \rangle_E$ the average quantity at $E$, and $Z(T)$ the partition function for the system at temperature $T$. To avoid the numerical overflow, $P(E,T)$ is calculated via:

$$P(E,T) = \frac{1}{Z'} e^{\ln g(E) - E/k_B T - \max(\ln g(E))} \qquad (7)$$

$$Z'(T) = \sum_E e^{\ln g(E) - E/k_B T - \max(\ln g(E))} \qquad (8)$$

where $\max(\ln g(E))$ is the maximum value of the logarithm of the DOS function for each star polymer system. The transition behaviors of these single star polymers are described with the temperature-dependent thermodynamic functions. The detailed description and comparison about the thermodynamic properties of the linear and star polymers are shown in the following parts.

**Internal energy and specific heat**

In canonical ensemble, the internal energy functions $U(T)$ of polymer chains are calculated via:

$$U(T) = \langle E \rangle_T = \sum_E E \cdot P(E,T) \tag{9}$$

The phase transition is defined by the large change in the $U(T) \sim T$ curve, e.g. when the system is undergoing an exothermic discontinuous phase transition, huge amounts of internal energy will release to the environment and an abrupt cliff shape will appear on the curve. Therefore, the phase transitions can be identified as a transition peak in the specific heat function $C_V(T)$ calculated with

$$C_V(T) = \frac{\partial U(T)}{\partial T} = \frac{\langle E^2 \rangle_T - \langle E \rangle_T^2}{k_B T^2} \tag{10}$$

Figure 5 compares the specific heat function for the three types of polymers mentioned above. For the linear polymers, the two-stage transitions on the $C_V(T)/N \sim T$ curves are shown in Fig. 5-a. For relative long chains $(L_a \geq 10)$, the sharp freezing transition peak appears accompanying with a shoulder-like collapsing transition peak at the higher temperature. With the increasing of the polymerization $N$, the freezing transition peak becomes sharper, and the corresponding transition temperature shifts towards the high temperature direction. Similar to the cluster-like packing phenomenon in off-lattice model studied by Seaton *et al.*[20], the special packing stability for small star polymers with certain monomer number is also observed in our lattice polymer model (see supporting materials in Fig. s1), which, however, is not the focusing point in the present work.

In Fig. 5-b, like the linear polymers, the two-stage transitions also exist for the 6-arm star polymers. However, compared with the linear polymers (Fig. 5-a), the continuous coil-globule transition becomes weak especially for $L_a < 20$, which can hardly be identified. When the arm number further increases ($N_a=12$ in Fig. 5-c), the CG transition shoulder becomes much weaker. The reason is that, with the increasing of the arm number, the monomer density of star polymer increases. It weakens the collapse condition of the coil-globule transition. Since the star polymers have a special structure with a dense core and a loose outer shell, the CG transition may appear when the arm length is sufficiently long and the average monomer density is lower.

**Canonical scaling analysis**

The transitions from canonical analysis are extended to the thermodynamic limit using the scaling method, which can be used to describe the properties of an infinite chain. It is believed that, for a finite size polymer, the relationship between the liquid-crystal transition temperature $T_{tr}$ and the polymerization $N$ behaves as $N^{-1/3}$, which is mainly due to the surface/volume effect under the finite size condition: [57]

$$T_{tr}(N) = T^*_{freezing} - \frac{a}{N^{1/3}} \tag{11}$$

where the freezing transition temperature of an infinite chain $T^*_{freezing}$ is obtained at $N \to \infty$. All the freezing transition temperatures obtained from the specific heat functions are plotted in Fig. 6. All points converge to a common linear relationship towards $N \to \infty$, which is in good agreements with the scaling prediction. More importantly, it indicates that the scaling law of the liquid-crystal transition of a single polymer chain is not affected by the chain structure. When $N \to \infty$, the infinite chain freezing transition temperature is achieved $T^*_{freezing} = 2.13$, which equals the prediction from past study[34] (about $T^*_{freezing} = 2.18$) for linear polymer under statistical error.

From theoretical study[57], the coil-globule transition or the θ transition temperature shifts with the $N^{1/2}$ for linear polymer, which can be described as the viral expansion ($T_\theta = T^*_\theta - b_1 N^{-1/2} + b_2 N^{-1} + ...$). The θ temperature can be determined by the specific heat function or the common intersection point of the $<R^2_g>_T/N \sim T$ curves for different chain sizes. For star polymers, it is very difficult to locate the transition temperature accurately with both of the two methods since the CG transition is too weak compared with the LC transition. Instead of that, the structural transition analysis is performed to describe the CG transition.

**(3) Structural transition analysis**

The arms in a star polymer are connected by a common point, which confines some conformations of the star polymer. This special structure brings star polymers unique thermodynamic properties and rich phase behaviors[58-60]. In this work, the mean square radius of gyration ($R_g^2$) and hydrodynamic radius ($R_h$) are used to describe the transition behavior. The micro-canonical ensemble $<R_g^2>_E$ and $<R_h>_E$ are given[57] by Eq. 12:

$$\left\langle R_g^2 \right\rangle_E = \frac{1}{2N^2} \left\langle \sum_{i=1}^{N} \sum_{j=1}^{N} |\mathbf{r}_{ij}|^2 \right\rangle_E \quad \text{and} \quad \left\langle R_h \right\rangle_E^{-1} = \frac{1}{N^2} \left\langle \sum_{i=1}^{N} \sum_{j=1(i \neq j)}^{N} |\mathbf{r}_{ij}|^{-1} \right\rangle_E \qquad (12)$$

here, $\mathbf{r}_{ij}$ is the distance vector and the summation is implemented for all monomer pairs. $<R_g^2>_E$ and $<R_h>_E$ in the extended ensemble are calculated through a second stage sampling using WL type acceptance criterion (see Eq. 2) with the DOS function obtained from the first stage sampling. In the first stage for achieving DOS function, final updater is $\ln f = 10^{-7}$. This updater is still adopted in the second stage instead of using a fixed DOS function. Based on the dynamical nature of the WL algorithm, this small updater enables the sampler running across all range energy states to traverse the conformation space and it really makes a *"flat"* enough histogram. Compared with the fixed DOS version of sampling, which performed with a multi-canonical like sampling algorithm[23], the present method, using the DOS function as a template, makes the histogram flatten faster and more stable. In the second stage calculation, the statistical error in $\ln g(E)$ is $\sqrt{\ln f}$ and enough samples in each energy bin up to $10^8$ are collected to ensure reliable statistical precision. The results of the second stage sampling are shown in supporting information (Fig. s2).[61] The decreasing of the radius function with the $E$ or $T$ indicates the collapse of the whole polymer.

The ratio of $R_g$ and $R_h$, called shape factor $\rho$ [62,63], is used for analysis. The shape factor, which is widely used in light scattering techniques, can be applied to probe the shape, size or structure of a particle system or an aggregation. Here, it is used to measure the conformational changing and CG transition of star polymers. The definition of the micro-canonical shape factor $\langle \rho \rangle_E$ is

$$\langle \rho \rangle_E = \langle R_g / R_h \rangle_E \tag{13}$$

Experimentally, the $R_g$ and $R_h$ are obtained from the static and dynamic light scattering techniques and the shape factor $\rho$ can be calculated by their ratio. The relationship between the shape factor $\rho$ and its corresponding structure of the aggregation has been studying for many years. Past studies[63-66] showed that, for an ideal coil, $\rho = 1.5$, while for a solid sphere, $\rho = 0.77$. Figure 7 shows the shape factor as a function of the specific energy for the above three sets of samples.

From the shape factor function, one can observe the conformation at the high energy state has a larger $\rho$ than the one at the lower energy state. Especially, for the longest linear polymer at the highest energy state, $\rho$ is about 1.4, which is close to the value of the ideal coil. With the increasing of the arm number, the difference in $\rho$ between the highest and lowest energy state becomes small. There is an obvious valley region in all panels, while $<R^2_g>_E$ and $<R_h>_E$ functions also appear such region but unclear (see supporting information in Fig. s2).[61]

It is believed that the shape factor is more effective to describe the CG and LC transitions from the structural change. In the general theory of the micro-canonical thermodynamics[49], the first order phase transition is signaled as a "convex intruder"[49] in the entropy function. The entropy function here serves as a partition function in the micro-canonical ensemble analysis. The micro-canonical temperature can be derived as

$$t(E) = \left( \frac{\partial S(E)}{\partial E} \right)^{-1} \tag{14}$$

The inverse temperature function can be obtained by acting derivative of the entropy function on $E$. When a system is undergoing a first order transition, with the increasing of the system energy, its temperature, however, decreases.[23,32,49] A comparison of the inverse temperature with the shape factor as a function of the specific energy for three polymer samples is shown in Fig. 8. The minimal points of the valleys on the curves for the inverse temperature and shape factor are located. It is found that the specific energy value is exactly the same as that of the shape factor. This fact indicates that when the polymer is undergoing a first order transition, with the increasing of its energy, its shape factor and temperature decreases together. Therefore, this result shows the shape factor $\rho$ can serve as an indicator for phase transition. Considering the structural transition in canonical ensemble, the shape factor as a function of canonical temperature can be calculated via

$$\langle \rho \rangle_T = \sum_E \langle \rho \rangle_E P(E,T) \tag{15}$$

With the temperature decreasing, the polymer chain undergoes a collapse process. The step region appears on both of the $<R_{g,h}>_T \sim T$ and $<\rho>_T \sim T$ curves indicating the first order LC transition (see supporting information in Fig. s3 and Fig. s4).[61]

Similar to the specific heat function, the derivative of the shape factor function with temperature $T$: $d\langle \rho \rangle_T / dT$ is used to quantitatively locate the transition points in Fig. 9. Compared with the transition peak in $C_V$ as shown in Fig. 4, the shape factor curve shows a more apparent transition peak, especially for the coil-globule transition. It is more accurate to determine the position of transition peaks in Fig. 9. These curves clearly show that with the increasing of the arm number, the CG transition becomes weaker compared with the LC transition.

For linear polymers and 6-arm star polymers, the scaling behaviors of the two transitions are plotted in Fig. 10. It is shown that the freezing transition scaling of the star polymers does not depend on the chain structure and there is a single scaling value for LC transition. The limit value of the freezing transition temperature at $N \to \infty$ is at about $T^*_{freezing} = 2.15$ which is close to the prediction of 2.13 obtained by specific heat function. For the CG transition, the circumstance is more complex. With the increasing of the arm number, the CG transition shifts close to the LC transition temperature. As the description from Rampf $et\ al.$[34], the truncated virial expansion may not enough to describe this model. We also found that the second virial coefficient is needed to describe the scaling results. The fitting according to the virial expansion $T_\theta = T_\theta^* - b_1 N^{-1/2} + b_2 N^{-1}$ is applied in Fig. 10 and both of the CG transition data can be well described by this relation. At the thermodynamic limit, there is only one coil-crystal transition within this model. The dense core and loose shell structure of star polymers makes the globule and coil states combined in a star polymer itself, which can be varied with the growing of its arms. The differences in the chain structure between linear and star polymer only affect the coil-globule transition under the finite size condition. Since the star polymers process higher monomer density, the shifting trend of the CG transition towards the LC transition in Fig. 10 should because of their higher density chain, which requires a lower temperature for the attractive force to overcome the excluded volume effects.

## V. CONCLUSION

In this work, we investigated the influence of the polymer structures on the thermodynamic transitions of star polymers. Through the precise calculation of the thermodynamic quantities using a parallel version of the 1/$t$ Wang-Landau sampling algorithm, we proved that the 1st order freezing transitions of the star polymers were not affected by their structure, while the 2nd order coil globule transitions were indeed varied by the structural factor of star polymers (i.e. the arm number). It is found that, compared with the specific heat function, the shape factor i.e. the ratio of the radius of gyration and hydrodynamic radius, is more powerful to reveal the coil-globule transition. With the increasing of the arm number, the coil-globule transition gradually approaches to the freezing transition. Similar to the previous studies[21-23] about influence of the attractive range of the interaction, the linear polymer with short-range attractive potential also shows this shifting manner of the CG transition. However, the LC transition is also influenced when the attractive well changed. Our results clearly show that, by controlling the arm number and length of the star polymers, we can change the coil-globule transition to obtain desired thermodynamic properties. Although our results are obtained from star polymers, it is believed that similar rules also appear for pom-pom, comb or brush polymers.

**Supporting information**: The specific heat function for small star polymers, the relations of $<R_g>_E \sim E$, $<R_h>_E \sim E$, $<R_g>_T \sim T$, $<R_h>_T \sim T$, and the shape factor as a function of specific energy are shown in the supporting materials.[61]

## VI. ACKNOWLEDGMENT

The project is supported by NSFC (Nos. 20804028 and 20974078), KPCME (No. 109043) and SRFDP (No. 200800561006).

# FIGURE CAPTIONS:

| Fig. 1 | (Color online) Snapshots of a typical expanded coil ($E=0$) and corresponding collapsed crystallite ($E=-4N$) for two types of polymers, where upper panel is a linear polymer ($N_a=2$ and $L_a=240$), and the lower panel is a 12-arm star polymer ($N_a=12$ and $L_a=60$). The terminal monomer is colored with red, and linear chain is in yellow. Images are rendered by VMD.[67] |
|---|---|
| Fig. 2 | The Schematic diagram of the hybrid movement algorithm including four types of simple movements, i.e. BFM movement, pivot movement, crankshaft movement and center imaging movement. |
| Fig. 3 | (Color online) The entropy function $\ln g(E)$ (left panels) and the corresponding average entropy function $\ln g(E)/N$ (right panels) as a function of the specific energy $E/N$. The apparent intersection regions in the left panels are magnified in the inset chart. The blue arrows in the right panels label the shifting directions of the maximum entropy point with the increasing of the arm length $L_a$. a: Linear polymers; b: 6-arm star polymers; c: 12-arm star polymers. |
| Fig. 4 | The $\ln g(E)$ of the intersection regions in the $\ln g(E) \sim E/N$ curves as a function of $N_a^3$, which is fitted as a straight line. |
| Fig. 5 | (Color online) Dependence of the specific heat per monomer $C_V(T)/N$ on the canonical temperature $T$ (unit in $k_B$). a: Linear polymers; b: 6-arm star |
| | polymers; c: 12-arm star polymers. |
| Fig. 6 | Freezing transition temperature as a function of $N^{1/3}$, where all polymer samples show a common linear relation towards the direction of an infinite chain. The linear relation has been extrapolated to $N\rightarrow\infty$ to obtain the limit freezing temperature at about $T_{freezing}$=2.13. |
| Fig. 7 | (Color online) The shape factor $\rho$ as a function of the specific energy. The valley region in each panel is labeled with a blue arrow. a: Linear polymers; b: 6-arm star polymers; c: 12-arm star polymers. |
| Fig. 8 | (Color online) Comparison of the inverse temperature with the corresponding shape factor as a function of the specific energy, where the three polymer samples are from the three sets of polymer chains mentioned above. The valley positions in both panels are labeled with three vertical lines showing that they have the identical energy states. |
| Fig. 9 | (Color online) The derivative of the shape factor $\rho$ as a function of the canonical temperature (unit in $k_B$). a: Linear polymers; b: 6-arm star polymers; c: 12-arm star polymers. |
| Fig. 10 | (Color online) The LC and CG transition temperatures as a function of $N^{-1/2}$ for the linear polymers and the 6-arm star polymers respectively. The two transitions are fitted according to the scaling equation shown in the chart. The symbols are derived from Fig. 9. See the main text for a discussion. |

| | Coil | Crystallite |
|---|---|---|
| | 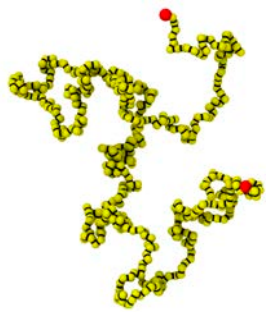 | 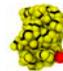 |
| | 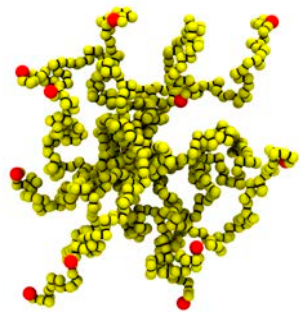 | 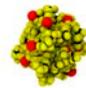 |

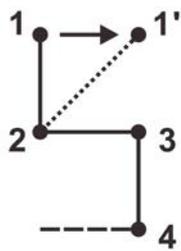
**BFM**

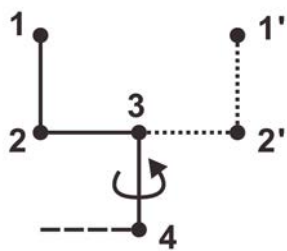
**Pivot**

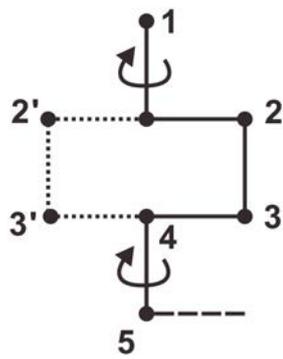
**Crankshaft**

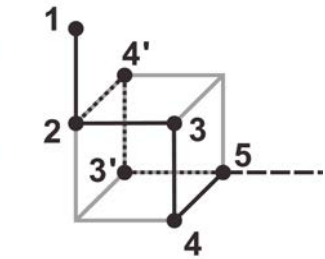
**Center imaging**

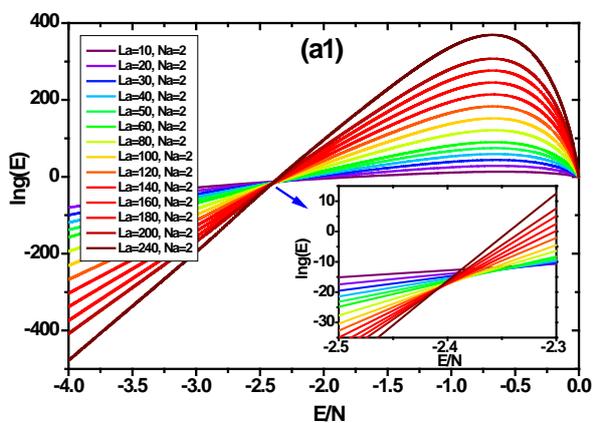
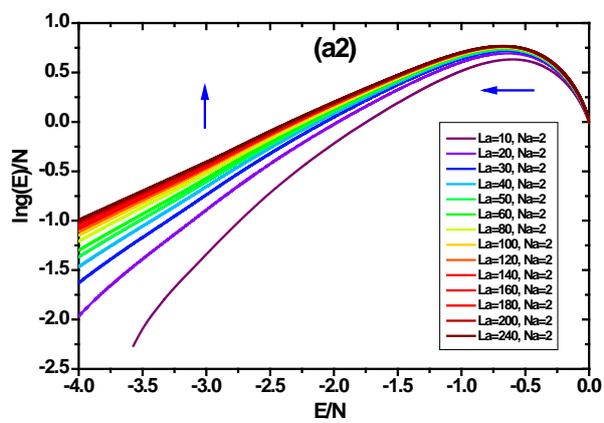
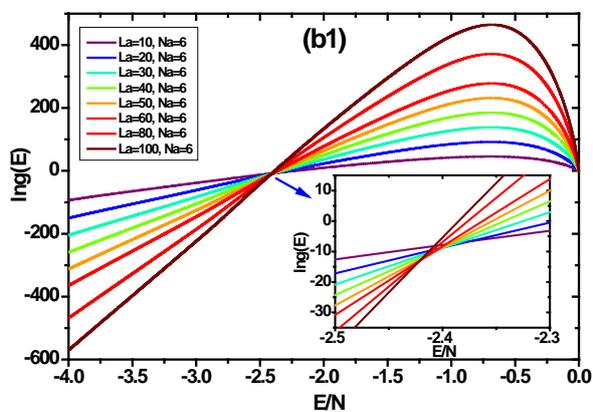
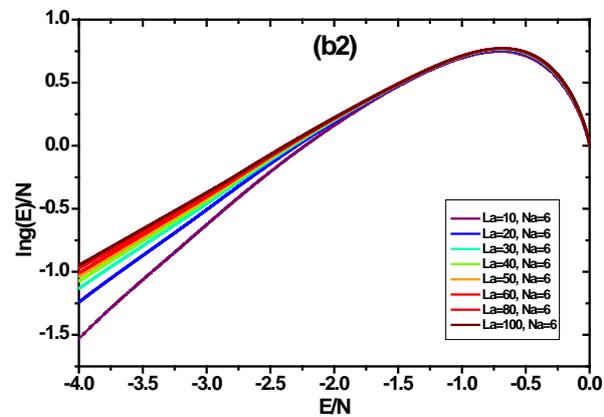
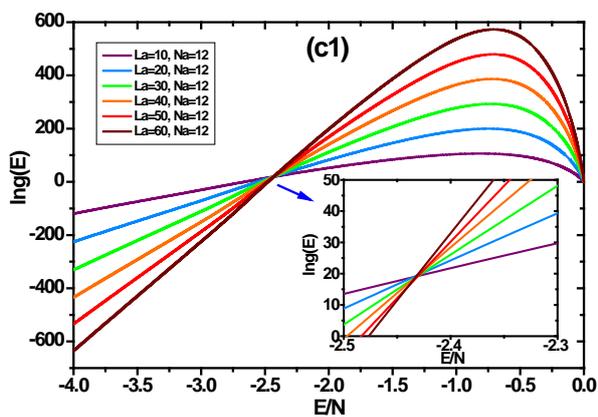
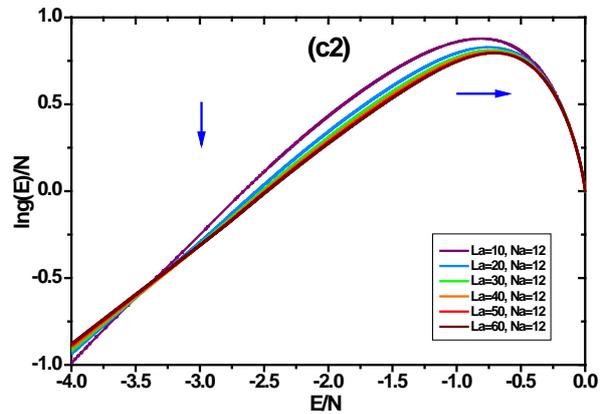

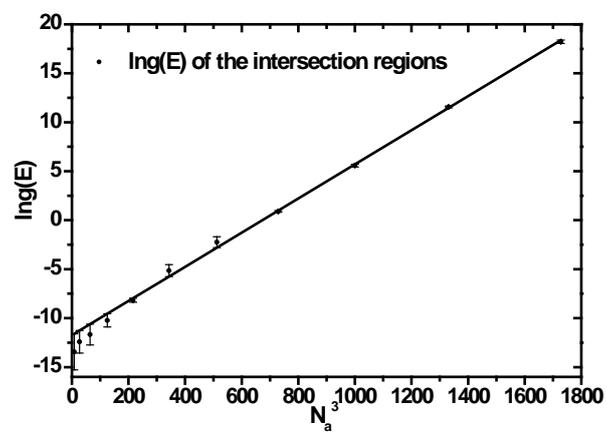

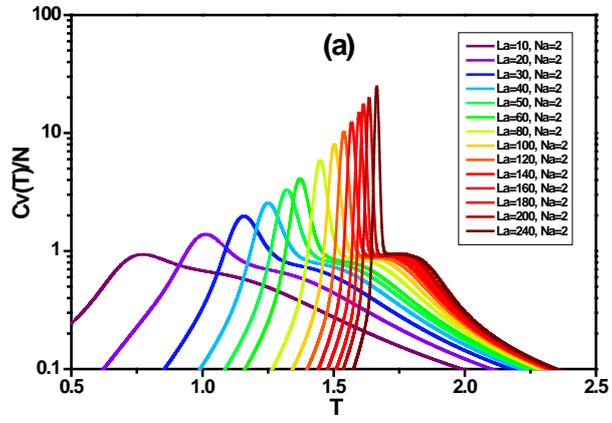

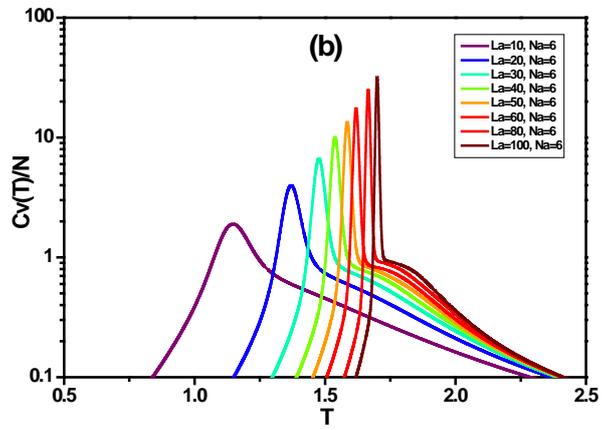

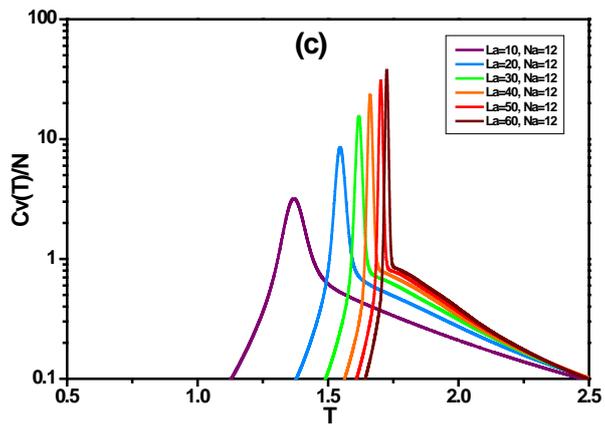

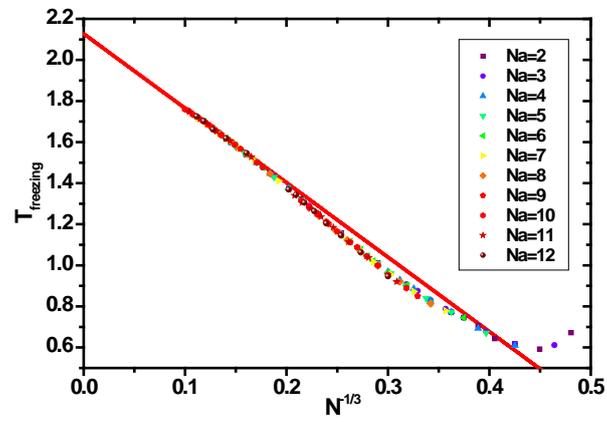

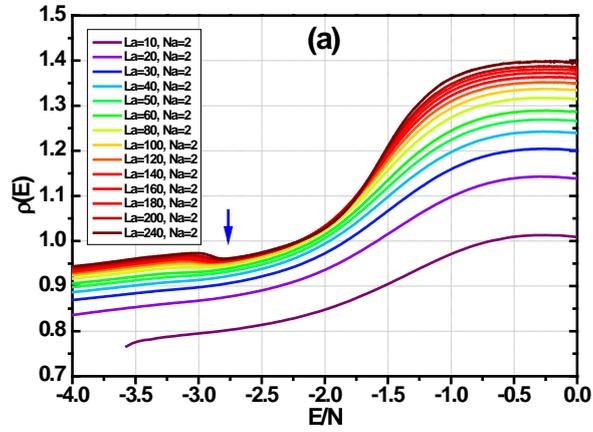

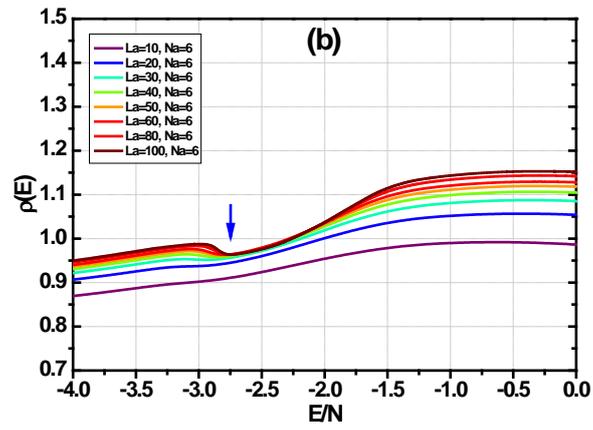

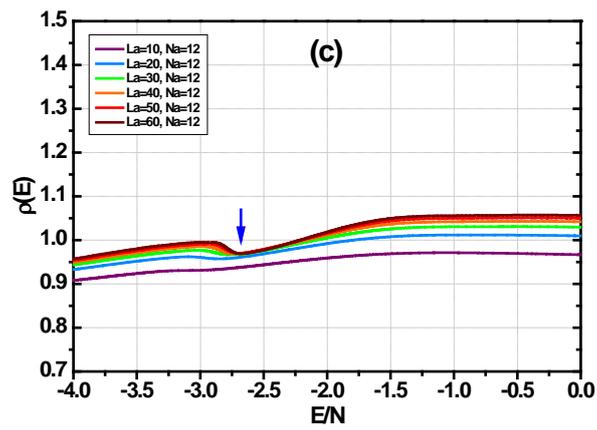

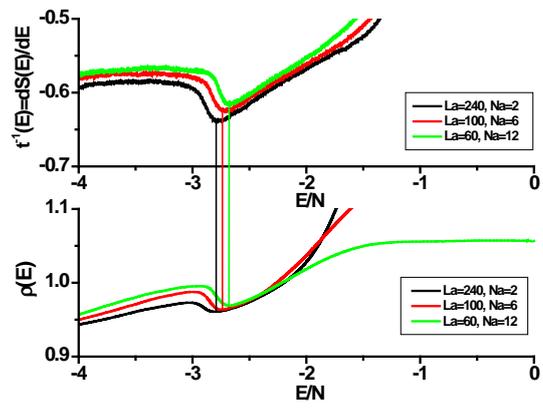

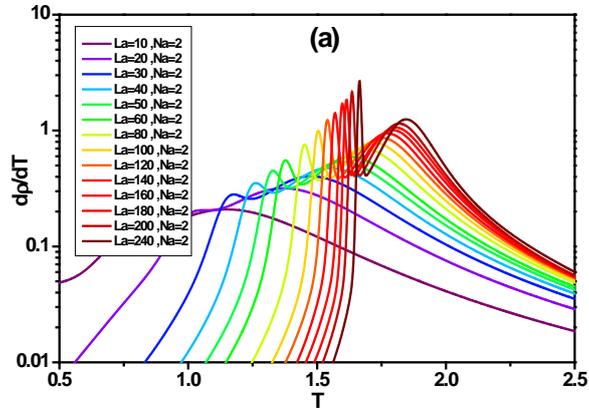

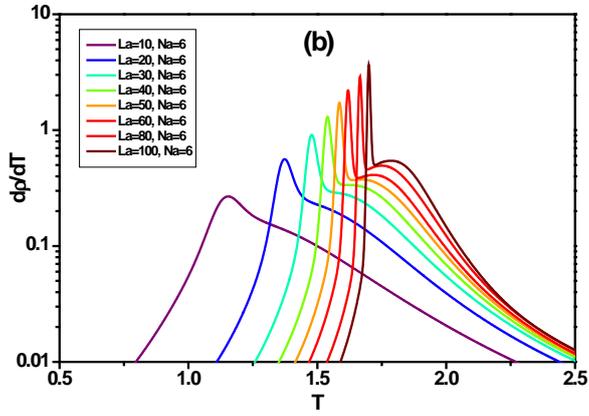

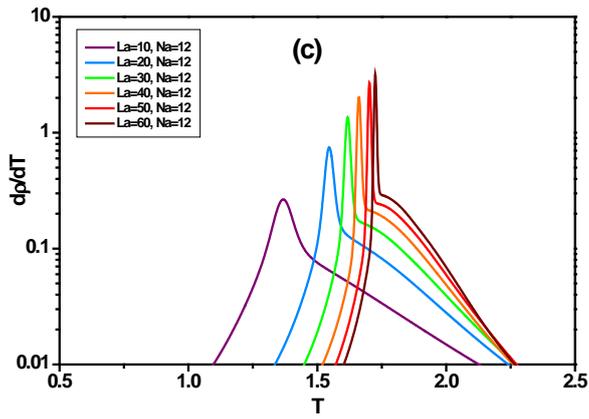

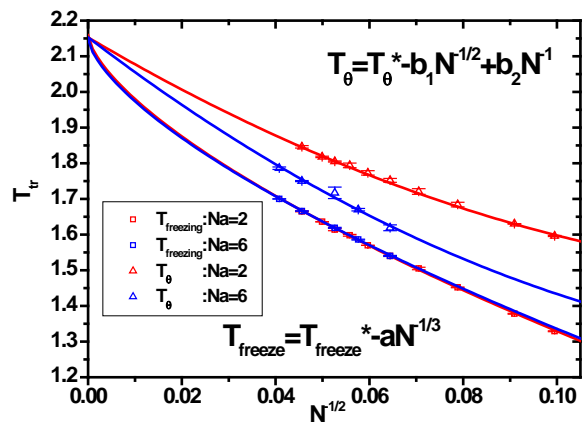